\documentclass[10pt,conference,romanappendices]{IEEEtran}
\usepackage[cmex10]{amsmath}
\usepackage[hidelinks]{hyperref}
\usepackage{xspace}
\usepackage{wrapfig}
\usepackage{graphicx}
\usepackage{xcolor}
\usepackage{soul}

\newtheorem{example}{Example}
\newtheorem{definition}{Definition} 
\newtheorem{proposition}{Proposition}

\newtheorem{remark}{Remark}

\newenvironment{mmproof}{\hspace{8pt}\ti{Proof:}}{}

\newcommand{\Sr}{\mbox{$\mi{Search}(F,\pnt{p},r)$}\xspace}
\newcommand{\sr}{\mbox{$\mi{Search}$}\xspace}
\newcommand{\sd}{\ti{Schn\_der}\xspace}
\newcommand{\spc}{$B^{|X|}$\xspace}
\newcommand{\Gs}{\mbox{$\mi{Gen\_SSC}$}\xspace}
\newcommand{\gs}{\mbox{$\mi{Gen\_SSP}$}\xspace}
\newcommand{\Nb}[2]{\mbox{$\mathit{Nbhd}(\pnt{#1},#2)$}\xspace}
\newcommand{\nb}[2]{\mbox{$\mathit{Nbhd}(#1,#2)$}\xspace}
\newcommand{\ie}{\emph{i.e.}, }
\newcommand{\eg}{\emph{e.g.}, }
\newcommand{\pnt}[1]{{\mbox{\boldmath $#1$}}}

\newcommand{\V}[1]{\mbox{$\mathit{Vars}(#1)$}}

\newcommand{\s}[1]{\mbox{$\{#1\}$}}

\newcommand{\nGz}[2]{$G_{non-\{z\}}$}

\newcommand{\prr}[1]{\mi{Prev}(\boldsymbol{q})}

\newcommand{\mi}[1]{\mathit{#1}}
\newcommand{\ti}[1]{\textit{#1}}
\newcommand{\tb}[1]{\textbf{#1}}

\newcommand{\ttt}{\>\>\>}
\newcommand{\tttt}{\>\>\>\>}

\newcommand{\Tt}{\>\>}

\newcommand{\Comment}[1]{}

\newcommand{\bm}[1]{{\boldmath $#1$}}

\begin{document}

\title{Solving SAT By Computing A Stable Set Of Points In Clusters}
\author{\IEEEauthorblockN{Eugene Goldberg}
 \IEEEauthorblockA{
email: eu.goldberg@gmai.com}}

\maketitle
\begin{abstract}
Earlier we introduced the notion of a stable set of points (SSP). We
proved that a CNF formula is unsatisfiable iff there is a set of
points (i.e. complete assignments) that is stable with respect to this
formula. Experiments showed that SSPs for CNF formulas of practical
interest are very large. So computing an SSP for a CNF formula point
by point is, in general, infeasible. In this
report\footnote{The idea of computing SSPs in clusters was first presented in the
technical report~\cite{old_clusters}. Since the latter is only
available on the author's website we decided to publish a new version
to make it more accessible. In this publication we introduced many
changes.  In particular, we changed the procedure \Gs where clusters
are represented by cubes to make it more practical. We also presented
an example of how \Gs works. Besides, we added a discussion of how
computing SSPs in clusters benefits parallel SAT solving.

}, we show how an SSP can be computed
in “clusters”, each cluster being a large set of points that are
processed “simultaneously”. The appeal of computing SSPs is
twofold. First, it allows one to better take into account formula
structure and hence, arguably, design more efficient SAT algorithms.
Second, SAT solving by SSPs facilitates parallel computing.
\end{abstract}

\section{Introduction}
In~\cite{ssp,annals}, we introduced the notion of a \tb{stable set of
  points (SSP)} for CNF formulas. (By points here we mean complete
assignments.)  We showed that to prove $F$ unsatisfiable it suffices
to construct an SSP for $F$. If $F$ is satisfiable, no SSP exists.
The appeal of SSPs is twofold. First, they are formula specific, which
allows one to exploit formula structure (e.g. formula symmetries).
Second, an SSP can be viewed as a proof of unsatisfiability where
different parts of this proof are related weakly. This facilitates
parallel computing.

Even though for some classes of formulas there are polynomial size
SSPs, in general, SSPs are exponential in formula size. A simple
procedure for building an SSP ``point by point'' was given
in~\cite{ssp}.  Experiments showed that the number of points in an SSP
grew very large even for small CNF formulas. This implies that
building an SSP point by point is, in general, impractical. To address
this problem, it was suggested in~\cite{ssp,annals} to compute an SSP
“in clusters” thus processing many points simultaneously. In this
report, we describe computing SSPs in clusters in greater detail.

The contribution of this report is fourfold. First, we introduce the
notion of a \tb{stable set of clusters} (\tb{SSC}). The latter
represents an SSP implicitly and can be computed much more
efficiently.  Although we introduce only the notion of clusters
consisting of points, the stability of more complex objects (like
clusters of clusters of points) can be studied.  Second, we describe
how the notion of an SSC works in testing the satisfiability of
symmetric formulas (in particular, pigeon-hole formulas). Third, we
show how one can build an SSC where clusters are specified by
cubes. Fourth, we argue that computing an SSC facilitates parallel SAT
solving.

This report is structured as follows. Section~\ref{sec:defs} recalls
the notion of SSPs and gives relevant definitions.  In
Section~\ref{sec:clusters}, we introduce the notion of a stable set of
clusters.  Section~\ref{sec:symm} describes how a stable set of
clusters is computed for symmetric formulas.  In
Section~\ref{sec:cubes} we present \Gs, a procedure for computing a
stable set of clusters where clusters are cubes.  A discussion of \Gs
is presented in
Section~\ref{sec:discussion}. Sections~\ref{sec:background}
and~\ref{sec:conclusions} provide some background and conclusions.

\section{Recalling Stable Sets Of Points}
\label{sec:defs}
In this section, we recall the notion of SSP introduced in~\cite{ssp}
and give relevant definitions.

\subsection{Definitions}
\label{ssec:defs}

\begin{definition}
Denote by \bm{B} the set \s{0,1} of values taken by a Boolean
variable. Let $X$ be a set of Boolean variables.  An \tb{assignment}
to $X$ is a mapping $X' \mapsto B$ where $X' \subseteq X$.  If $X' =
X$ this assignment is called \tb{a complete one}. We will denote by
\bm{B^{|X|}} the set of complete assignments to $X$. A complete
assignment to the variables of $X$ is also called a \tb{point} of
$B^{|X|}$.
\end{definition}

\begin{definition}
A \tb{literal} of a Boolean variable $x$ is either $x$ itself or its
negation.  A disjunction of literals is called a \tb{clause}. A
formula that is a conjunction of clauses is said to be in the
conjunctive normal form (\tb{CNF}). A clause $C$ is called
\tb{satisfied} by an assignment \pnt{p} if $C(\pnt{p}) = 1$.
Otherwise, the clause $C$ is called \tb{falsified} by \pnt{p}. The
same applies to a CNF formula and an assignment \pnt{p}.
\end{definition}

\begin{definition}
Let $F$ be a CNF formula. The \tb{satisfiability problem} (SAT for
short) is to find an assignment satisfying all the clauses of
$F$. This assignment is called a \tb{satisfying assignment}.
\end{definition}

\begin{definition}
\label{def:nbhd}
Let $\pnt{p} \in B^{|X|}$ be a point (\ie a complete assignment to
$X$) falsifying a clause $C$. The \tb{1-neighborhood of} \pnt{p} with
respect to $C$ (written \bm{\Nb{p}{C}}) is the set of points satisfying $C$
that are at Hamming distance 1 from \pnt{p}.
\end{definition}

It is not hard to see that the number of points in \Nb{p}{C} equals
that of literals in $C$.
\begin{example}
Let $C=x_1 \vee \overline{x}_3 \vee x_4$ be a clause specified in the
Boolean space of 4 variables $x_1\dots,x_4$. Let\linebreak
$\pnt{p}=(x_1=0,x_2=1,x_3=1,x_4=0)$ be a point falsifying $C$.  Then
\Nb{p}{C} consists of the following three points:
\pnt{p_1}=({\boldmath $x_1=1$}$,\,x_2=1,x_3=1,x_4=0$),
$\pnt{p_2}=(x_1\!=\!0,\,x_2=1$,\,\bm{x_3=0},\,$ x_4=0$),
$\pnt{p_3}=(x_1=0,x_2=1, x_3=1,$ \bm{x_4=1}). Points \pnt{p_1},
\pnt{p_2}, \pnt{p_3} are obtained from \pnt{p} by flipping the value
of variables $x_1$, $x_3$, $x_4$ respectively.
\end{example}

\begin{definition}
Given a formula $F$, denote by {\boldmath \V{F}} the set of its
variables. Denote by \bm{Z(F)} the set of complete assignments to
\V{F} falsifying \bm{F}. If $F$ is unsatisfiable, $Z(F) = B^{|X|}$
where $X= \V{F}$.
\end{definition}

\begin{definition} Let $F$ be
a CNF formula and $P$ be a subset of the set of falsifying points
$Z(F)$. A function $g$ mapping $P$ to $F$ is called a \tb{transport
  function} if, for every $\pnt{p} \in P$, the clause $g(\pnt{p})$ is
falsified by \pnt{p}. In other words, a transport function $g:P
\mapsto F$ is meant to assign each point $\pnt{p} \in P$ a clause of
$F$ that is falsified by \pnt{p}.  We call the mapping $P \mapsto F$
above a transport function because it allows one to introduce some
kind of “movement” of points in the Boolean space.
\end{definition}

\begin{definition}
\label{def:ssp}  
Let $P$ be a nonempty subset of $Z(F)$ where $F$ is a CNF formula. The
set $P$ is called \tb{stable} with respect to $F$ and a transport
function $g: P\mapsto F$, if $\forall \pnt{p} \in P$,
$\Nb{p}{g(\pnt{p})} \subseteq P$.  Henceforth, if we just say that a
set of points $P$ is stable with respect to a CNF formula $F$, we mean
that there is a transport function $g:P \mapsto F$ such that $P$ is
stable with respect to $F$ and $g$.
\end{definition}

\begin{example}
Consider an unsatisfiable CNF formula $F$ consisting of 7 clauses:
$C_1=x_1 \vee x_2$, $C_2=\overline{x}_2\vee x_3$,
$C_3=\overline{x}_3\vee x_4$, $C_4=\overline{x}_4\vee x_1$,
$C_5=\overline{x}_1\vee x_5$, $C_6=\overline{x}_5\vee x_6$,
$C_7=\overline{x}_6\vee \overline{x}_1$. Clauses of $F$ are composed
of the six variables $x_1,\dots,x_6$. Let $P = \s{p_1,\dots,p_{14}}$
where $\pnt{p_1}=000000, \pnt{p_2}=010000, \pnt{p_3}=011000,
\pnt{p_4}=011100, \pnt{p_5}=111100,\pnt{p_6}=111110, \pnt{p_7}=111111,
\pnt{p_8}=011111, \pnt{p_9}=011011, \pnt{p_{10}}=010011,
\pnt{p_{11}}=000011, \pnt{p_{12}}=100011, \pnt{p_{13}}=100010,
\pnt{p_{14}}=100000$.  (Values of variables are specified in the order
variables are numbered. For example, \pnt{p_4}=
($x_1\!=\!0,x_2\!=\!1,x_3\!=\!1,x_4\!=\!1,x_5=0,x_6=0$).  The set $P$
is stable with respect to the transport function $g$ specified as:
$g(\pnt{p_1})\!=\!C_1,\:
g(\pnt{p_2})\!=\!C_2,\:g(\pnt{p_3})\!=\!C_3,\linebreak
g(\pnt{p_4})\!=\!C_4,\: g(\pnt{p_5})\!=\!C_5,\:
g(\pnt{p_6})\!=\!C_6,\: g(\pnt{p_7})\!=\!C_7,\linebreak
g(\pnt{p_8})\!=\!C_4,\: g(\pnt{p_9})\!=\!C_3,\:
g(\pnt{p_{10}})\!=\!C_2,\: g(\pnt{p_{11}})\!=\!C_1,\linebreak
g(\pnt{p_{12}})\!=\!C_7,\: g(\pnt{p_{13}})\!=\!C_6,\:
g(\pnt{p_{14}})\!=\!C_5$. It is not hard to see that $g$ indeed is a
transport function i.e. for any point \pnt{p_i} of $P$ it is true that
$C(\pnt{p_i})=0$ where $C=g(\pnt{p_i})$. Besides, every point
\pnt{p_i} of $P$ satisfies the condition $\Nb{p}{g(\pnt{p})} \subseteq
P$ of Definition~\ref{def:ssp}.  Consider, for example, point
$p_{10}=010011$. The value of $g(\pnt{p_{10}})$ is $C_2$ where
$C_2=\overline{x}_2\vee x_3$. The value of \Nb{p_{10}}{C_2} is
\s{\pnt{p_{11}}=000011, \pnt{p_9}=011011}. So, the latter is a subset
of $P$.
\end{example}

\begin{proposition}.
\label{prop:ssp}
If there is a set of points that is stable with
respect to a CNF formula $F$, then $F$ is unsatisfiable. 
\end{proposition}

The proof of this proposition is given in~\cite{annals}. The reverse
of Proposition~\ref{prop:ssp} is true too \ie for every unsatisfiable
formula $F$ there is an SSP. A trivial SSP is \spc where $X =
|\V{F}|$.

\subsection{Procedure For Building SSP}
\label{ssec:bld_ssp}
In this subsection, we recall a simple procedure introduced
in~\cite{ssp,annals} that generates an SSP point by point. We will
refer to it as \gs. The pseudocode of \gs is shown in
Figure~\ref{fig:ssp}. \gs accepts a CNF formula $F$ and returns either
a satisfying assignment or an SSP proving $F$ unsatisfiable. \gs
maintains two sets of points: $\mi{Boundary}$ and $\mi{Body}$. The set
$\mi{Boundary}$ (respectively $\mi{Body}$) consists of the reached
points whose neighborhood points have not been generated yet
(respectively are already generated).

%
%
\vspace{4pt}
\setlength{\intextsep}{4pt}
\setlength{\textfloatsep}{4pt}
\begin{figure}[h!]
\centering
\small
\begin{tabbing}
aaa\=bb\=cc\= dd\= \kill
$\gs(F)$ \{ \\
\scriptsize{1}\> $\pnt{p}_{init} := \mi{gen\_point}(F)$ \\
\scriptsize{2}\> $\mi{Body} =\emptyset, \mi{Boundary} = \s{\pnt{p}_{init}}$\\
\scriptsize{3}\> while ($\mi{Boundary} \neq \emptyset$) \{\\
\scriptsize{4}\Tt $\pnt{p} := \mi{pick\_next\_point}(\mi{Boundary})$\\
\scriptsize{5}\Tt $\mi{Boundary} := \mi{Boundary} \setminus \s{\pnt{p}}$\\
\scriptsize{6}\Tt $\mi{Body} := \mi{Body} \cup \s{\pnt{p}}$\\
\scriptsize{7}\Tt $H = \mi{falsified\_clauses}(F,\pnt{p})$\\
\scriptsize{8}\Tt if ($H = \emptyset$) return($\pnt{p},\emptyset$) // \pnt{p} is a satisf. assign. \\
\scriptsize{9}\Tt $C := \mi{pick\_clause}(H)$~~~// $g(\pnt{p}) := C$ \\
\scriptsize{10}\Tt $\mi{NewPnts}\!:=\!\Nb{p}{C}\!\setminus (\mi{Body} \cup \mi{Boundary})$\\
\scriptsize{11}\Tt $\mi{Boundary} := \mi{Boundary} \cup \mi{NewPnts}$\} \\
\scriptsize{12}\> return($\mi{nil},\mi{Body}$) // $\mi{Body}$ is an SSP, since $\mi{Boundary}=\emptyset$\\
\end{tabbing}
\vspace{-15pt}
\caption{Generation of SSP}
\label{fig:ssp}
\end{figure}

The set $\mi{Boundary}$ is initialized with a point $\pnt{p}_{init}$
while $\mi{Body}$ is originally empty (lines 1-2).  Then, in a
\ti{while} loop (lines 3-11), \gs does the following. It picks a point
\pnt{p} of $\mi{Boundary}$ to explore its neighborhood, removes
\pnt{p} from $\mi{Boundary}$ and adds it to $\mi{Body}$ (lines
4-6). Then it computes the set $H$ of clauses of $F$ falsified by
\pnt{p}. If $H$ is empty, \pnt{p} is a satisfying assignment. So, \gs
returns $\pnt{p}$ and an empty set indicating that no SSP is built
(line 8). Otherwise, a clause $C \in H$ is picked as the value of the
transport function $g$ at \pnt{p} in line 9 (\gs builds $g$ on the
fly). The points of \Nb{p}{C} that are not in $\mi{Body}$ yet and not
already in $\mi{Boundary}$ are added to $\mi{Boundary}$ (lines 10-11).

If the set $\mi{Boundary}$ is empty, it means that for every point
$\pnt{p} \in \mi{Body}$, the property $\Nb{p}{g(\pnt{p})} \subseteq
\mi{Body}$ holds. So, $\mi{Body}$ is an SSP and hence $F$ is
unsatisfiable (line 12).

\section{Computing A Stable Set Of Clusters}
\label{sec:clusters}
In this section, we introduce the notion of a stable set of clusters
of points. As we mentioned earlier, experiments show that computing an
SSP point by point is impractical. Building a stable set of clusters
can be viewed as a way to speed up SSP computing by processing many
points at once.

\begin{definition}
Let $F$ be a CNF formula and $P$ be a subset of $Z(F)$. Let $g$ be a
transport function $Z(F) \mapsto F$.  Denote by \bm{\mi{Nbhd}(P, g)}
the union of sets \Nb{p}{g(\pnt{p})}, $\pnt{p} \in P$ for all the
points of $P$. In other words, $\mi{Nbhd}(P, g)$ is the union of the
1-neighborhoods for all points $\pnt{p} \in P$ where
\Nb{p}{g(\pnt{p})} is computed with respect to the clause
$g(\pnt{p})$.
\end{definition}

\begin{definition}
Let $F$ be a CNF formula and $P_1,\dots,P_k$ be subsets of $Z(F)$. Let
$g_i$ be a transport function $P_i \mapsto F$, $i=1,\dots,k$. Suppose
that for every $P_i, i=1,..,k$ the property $\mi{Nbhd}(P_i,
g_i)\!\subseteq\!P_1\cup \dots\cup P_k$ holds. Then the set
\s{P_1,\dots,P_k} will be called a \tb{stable set of clusters (SSC)}
with respect to $F$ and transport functions $g_1,\dots,g_k$. (Here we
refer to a subset $P_i$ as a \tb{cluster} of points.)
\end{definition}

\begin{proposition}
Let $F$ be a CNF formula and $P_1,\dots,P_k$ be a stable set of
clusters with respect to $F$ and transport functions $g_1,..,
g_k$. Then $P_1 \cup \dots \cup P_k$ is an SSP and so $F$ is
\ti{unsatisfiable}.
\end{proposition}

A \ti{proof} of this proposition is given in the appendix. The same
applies to all \ti{new} propositions introduced in this paper.

\begin{remark}
\label{rem:kolmog_clust}
Note that if $P$ is an SSP for $F$, any set of $k$ subsets $P_i
\subseteq P$ forms an SSC if $P_1 \cup \dots \cup P_k = P$. However we
are interested only in clusters that make computing an SSC
efficient. Intuitively, such efficiency can be achieved if every
cluster $P_i$ is formed from the points that are somehow related to
each other.  More specifically, every cluster is supposed to satisfy
the following two properties. First, $P_i$ has a short description
regardless of its size \eg if $P_i$ is exponential in $|\V{F}|$. (One
can think of $P_i$ as a set of low Kolmogorov complexity.)  Second,
the 1-neighborhood of $P_i$ with respect to the transport function
$g_i$ can be easily computed.
\end{remark}

The notion of SSC is important for a few reasons.  Suppose for the
unsatisfiable formulas of some class there is an SSC with a polynomial
number of clusters (in formula size).  Then one can have an efficient
procedure for testing the satisfiability of the formulas of this
class. We substantiate this idea by the example of pigeon-hole
formulas. Another reason for studying SSCs is that they expose a deep
relation between models and formulas. So, one can get a better
understanding of the existing SAT algorithms (e.g. those based on
clause learning).

In this report, we consider only a two-level “hierarchy” of clusters,
namely, clusters consisting of points. However, one can introduce more
complex hierarchies (\eg clusters of clusters of points and so on).

\section{Testing Satisfiability Of Symmetric Formulas}
\label{sec:symm}
In this section, we show the relation between permutational symmetries
of a formula and its SSCs. In Subsection~\ref{ssec:symm_and_ssp}, we
recall the previous results on SSPs for symmetric formulas.
Subsection~\ref{ssec:symm_and_ssc} shows that the procedure for
solving symmetric formulas introduced in~\cite{annals} can be actually
interpreted as building an SSC. Finally, in Subsection~\ref{ssec:ph},
we consider SSCs for pigeon-hole formulas.

\subsection{Stable sets of points for symmetric formulas}
\label{ssec:symm_and_ssp}
\begin{definition}
Let $X$ be a set of Boolean variables. A \tb{permutation} \bm{\pi}
defined on the set $X$ is a bijective mapping of $X$ onto itself.
\end{definition}

\begin{definition}
Let $X=\s{x_1,\dots, x_n}$ be a set of Boolean variables.  Let
$\pnt{p}=(x_1,\dots, x_n)$ be a point of \spc . Let $\pi$ be a
permutation of $X$.  \tb{Denote by} \bm{\pi(\pnt{p})} the point
$(\pi(x_1),\dots,\pi(x_n))$.
\end{definition}

\begin{definition}
Let $F=\s{C_1,\dots,C_k}$ be a CNF formula. Let $\pi$ be a permutation
of \V{F}. Denote by $\pi(C_i)$ the clause obtained from $C_i$ by
replacing each variable\linebreak $x_m \in \V{C_i}$ with the variable
$\pi(x_m)$.  \tb{Denote by} \bm{\pi(F)} the set of clauses
\s{\pi(C_1),\dots,\pi(C_k)}
\end{definition}

\begin{definition}
Let $F$ be a CNF formula and $\pi$ be a permutation of \V{F}. Formula
$F$ is called \tb{symmetric} with respect to $\pi$ if $\pi(F)$
consists of the same clauses as $F$ (\ie each clause $\pi(C_i)$ of
$\pi(F)$ is identical to a clause $C_m$ of $F$).
\end{definition}

\begin{definition}
Let $X$ be a set of Boolean variables and $G$ be a group of
permutations of $X$. \tb{Denote by} \bm{\mi{symm}(\pnt{p},\pnt{p'},
  G)} the following binary relation between points of \spc. A pair of
points $(\pnt{p},\pnt{p'})$ is in $\mi{symm}(\pnt{p},\pnt{p'},G)$ iff
there is $\pi \in G$ such that $p' = \pi(\pnt{p})$.The relation
$\mi{symm}(\pnt{p},\pnt{p'},G)$ is an equivalence one and so it breaks
\spc into equivalence classes.
\end{definition}

\begin{definition}
Points \pnt{p} and \pnt{p'} of \spc are called \tb{symmetric} with
respect to a group $G$ of permutations of $X$ if they are in the same
equivalence class of $\mi{symm}(\pnt{p},\pnt{p'},G)$.
\end{definition}

\begin{proposition}
\label{prop:symm_hood}
Let $X$ be a set of Boolean variables and \pnt{p} be a point of
\spc. Let $C$ be a clause falsified by \pnt{p}. Let $\pi$ be a
permutation of $X$. Then
for each point \pnt{p'} of $\mi{Nbhd}(\pnt{p},C)$ there is a point
$\pi(\pnt{p'})$ of $\mi{Nbhd}(\pi(\pnt{p}),\pi(C))$.)
\end{proposition}

The proof is given in~\cite{annals}.

\begin{definition}
Let $F$ be a CNF formula that is symmetric with respect to a group $G$
of permutations of $X=\V{F}$. Let $P$ be a set of points of \spc
falsifying $F$. The set $P$ is called \tb{stable modulo symmetry} $G$
with respect to $F$ and a transport function $g: P \mapsto F$ if for
each point $\pnt{p} \in P$, every point \pnt{p'} of
$\mi{Nbhd}(\pnt{p}, g(\pnt{p}))$ is either in $P$ or there is a point
\pnt{p''} of $P$ that is symmetric to \pnt{p'}.
\end{definition}

\begin{proposition}
\label{prop:symm_ssp}
Let $F$ be a CNF formula, $P$ be a set of points of \spc , $X=\V{F}$,
that falsify $F$. Let $g: P \mapsto F$ be a transport function. If $P$
is stable modulo symmetry $G$ with respect to $F$ and $g$, then $F$ is
\tb{unsatisfiable}.
\end{proposition}

The proof is given in~\cite{annals}.

\subsection{Stable sets of clusters for symmetric formulas}
\label{ssec:symm_and_ssc}
Proposition~\ref{prop:symm_ssp} is proved in~\cite{annals} via
extending the set of points $P$ by adding the points symmetric to
those of $P$. The transportation function $g$ is also extended as
follows.  If $\pnt{p} \in P$ and $\pnt{p'} =\pi(\pnt{p})$, then
$g(\pnt{p'})$ is equal to $\pi(g(\pnt{p}))$ (In other words, for
symmetric points, the extended transport function $g$ assigns
symmetric clauses.) It is shown in~\cite{annals} that this extended
set of points is actually an SSP for $F$ with respect to the extended
transport function $g$.

Importantly, one can give a different interpretation of the extension
of $P$ above. Let $P=\s{\pnt{p_1},\dots, \pnt{p_s}}$. Let
$E(\pnt{p_i})$ be the equivalence class of the symmetry relation
$\mi{symm}(\pnt{p},\pnt{p'}, G)$ consisting of the points of \spc that
are symmetric to \pnt{p_i}. Then the set of clusters
$E(\pnt{p_1}),\dots,E(\pnt{p_s})$ \tb{form an SSC} because
$E(\pnt{p_1}) \cup \dots \cup E(\pnt{p_s})$ is exactly the extended
set of points described above and so this set is stable. (Note that if
points \pnt{p_i} and \pnt{p_j} of $P$ are symmetric, then
$E(\pnt{p_i})=E(\pnt{p_j})$.)

Sets $E(\pnt{p_i})$ meet the two requirements to clusters specified by
Remark~\ref{rem:kolmog_clust} of Section~\ref{sec:clusters}. On one
hand, each cluster is an equivalence class of the relation
$\mi{symm}(\pnt{p},\pnt{p'},G)$ and so the set of points of
$E(\pnt{p_i})$ can be easily described. On the other hand, the set
$\mi{Nbhd}(E(\pnt{p_i}),g)$ (where $g$ is the transport function
extended from the original function $P \mapsto F$ as described before)
is easy to compute. According to Proposition~\ref{prop:symm_hood},
$\mi{Nhbd}(\pi(\pnt{p}),\pi(C))$ and $\mi{Nbhd}(\pnt{p},C)$ consist of
points symmetric under $\pi$. Let $Nbhd(\pnt{p_i},C) =
\s{\pnt{p_{i_1}},\dots, \pnt{p_{i_m}}}$ (here $C$ is the clause
$g(\pnt{p_i})$). Then $\mi{Nbhd}(E(\pnt{p_i}),g)\!=\!E(\pnt{p_{i_1}})
\cup \dots \cup E(\pnt{p_{i_m}})$.

The procedure for building an SSC for a CNF formula $F$ with symmetry
$G$ is essentially identical to the procedure of~\cite{annals} for
building a set $P$ that is stable with respect to $F$ modulo symmetry
$G$. In turn, the procedure of~\cite{annals} is different from the one
shown in Figure~\ref{fig:ssp} only in one line of code (line
11). Namely, when building a set of points stable modulo symmetry $G$
this procedure does not add to \ti{Boundary} a point \pnt{p'} of
$\mi{Nbhd}(\pnt{p},C)$ if \ti{Total} contains a point that is
symmetric to \pnt{p'}.  Eventually this procedure builds a set of
points $P=\s{\pnt{p_1},\dots,\pnt{p_m}}$ that is stable with respect
to $F$ modulo symmetry $G$.

Importantly, one can interpret the procedure of~\cite{annals} as
building an SSC equal to \s{E(\pnt{p_1}),\dots,E(\pnt{p_m})}. This
procedure just uses points \pnt{p_i} of $P$ as representatives of
clusters $E(\pnt{p_i})$. Suppose, for instance, that a point \pnt{p'}
of $\mi{Nbhd}(\pnt{p},C)$ is not added to \ti{Boundary} because it is
symmetric to a point \pnt{p''} of \ti{Total}. In terms of SSCs this
just means that $E(\pnt{p'})=E(\pnt{p''})$ and so the cluster
$E(\pnt{p'})$ has already been ``visited''.

\subsection{Stable sets of clusters for pigeon-hole formulas}
\label{ssec:ph}
In this subsection, we illustrate the power of SSCs by the example of
pigeon-hole formulas. These are unsatisfiable CNF formulas that
describe the pigeon-hole principle. Namely, if $n > m$, then $n$
pigeons cannot be placed in $m$ holes so that no two pigeons occupy
the same hole. In~\cite{haken} A. Haken showed that pigeon-hole
formulas have only exponential size proofs in the resolution proof
system, which makes them hard for the SAT-solvers based on resolution.
Since the pigeon-hole principle is symmetric with respect to a
permutation of holes or pigeons, pigeon-hole formulas are highly
symmetric.

Let $\mi{PH}(n,m)$ denote a CNF formula encoding the pigeon-hole
principle above.  Let $G$ denote the permutational symmetry of
$\mi{PH}(n,m)$. In~\cite{annals} we showed that there is a set of
points $P = \s{\pnt{p_1},\dots,\pnt{p_{2m+1}}}$ that is stable for
$\mi{PH}(n,m)$ modulo symmetry $G$. Denote by $S(n,m)$ the union of
the equivalence classes $E(\pnt{p_i})$, $i=1,\dots,2m+1$ of the
relation $\mi{symm}(\pnt{p},\pnt{p'},G)$. The fact that $P$ is stable
modulo symmetry $G$ means that $S(n,m)$ is an SSP for
$\mi{PH}(n,m)$. On the other hand, this fact means that $\mi{PH}(n,m)$
has an SSC consisting of $2m+1$ clusters $E(\pnt{p_i})$. The size of
$E(\pnt{p_i})$ is exponential in $m$ and hence $S(n,m)$ is exponential
in $m$ too.  However, the size of the SSC above in terms of clusters
is \tb{linear} in $m$.

\section{Computing SSCs Using Cubes As Clusters}
\label{sec:cubes}
In this section, we introduce \Gs, a SAT procedure that builds a
special class of SSCs where clusters are cubes.
Subsections~\ref{ssec:more_defs} and~\ref{ssec:example} provide some
definitions and an example of how \Gs operates. In
Subsection~\ref{ssec:SSC_with_cubes}, we present the pseudocode of
\Gs.

\subsection{A few more definitions and examples}
\label{ssec:more_defs}
\begin{definition}
\label{def:cube}
Let $X =\s{x_1,\dots, x_n}$ be a set of Boolean variables. A \tb{cube}
$P$ of $B^{|X|}$ is a subset of $B^{|X|}$ that can be represented as
$B_1\times \dots \times B_n$, where $B_i$ is a non-empty subset of $B$
and $'\times'$ means the Cartesian product. The components $B_i$ equal
to \s{0} or \s{1} are called \tb{literal components} of $P$.
\end{definition}

\begin{definition}
We will say that a cube $P$ \tb{satisfies} (respectively
\tb{falsifies}) a clause $C$ if every point $\pnt{p} \in P$ satisfies
(respectively falsifies) $C$.
\end{definition}

\begin{definition}
Let $X =\s{x_1,\dots, x_n}$ be a set of Boolean variables.  Let $P =
B_1\times\dots\times B_n$ be a cube of $B^{|X|}$ and $B_i$ be equal to
\s{0,1}. Let $P',P''$ be the cubes obtained from $P$ by replacing the
set $B_i$ with sets \s{0} and \s{1} respectively. We will say that
cubes $P'$ and $P''$ are obtained from $P$ by \tb{splitting} on
the variable $x_i$.
\end{definition}

\begin{definition}
Let $X =\s{x_1,\dots, x_n}$ be a set of Boolean variables.  Let $C$ be
a clause, $\V{C} \subseteq X$. Denote by \bm{\mi{Unsat}(C)} the set of
all points of $B^{|X|}$ that falsify $C$. It is not hard to see that
$\mi{Unsat}(C)$ \tb{is a cube} of $B^{|X|}$.
\end{definition}

\begin{example}
Let $C=x_2\vee \overline{x}_4$ and $X=\s{x_1, x_2, x_3, x_4}$. Then
$\mi{Unsat}(C)$ equals
$\s{0,1}\times\s{0}\times\s{0,1}\times\s{1}$. In other words,
$\mi{Unsat}(C)$ consists of all the points of $B^{|X|}$ for which
$x_2=0$ and $x_4=1$. So, the second and fourth components of
$\mi{Unsat}(C)$ are literal components.
\end{example}

\begin{definition}
\label{def:nbhd_in_dir}
Let $X =\s{x_1,\dots, x_n}$ be a set of Boolean variables.  Let
\pnt{p} be a point of \spc. Denote by \bm{\Nb{p}{x_i}} the
neighborhood of \pnt{p} in direction $x_i$, i.e. the one-element set
\s{\pnt{p'}} where the point \pnt{p'} is obtained from \pnt{p} by
flipping the value of $x_i$ in \pnt{p}.
\end{definition}

From Definition~\ref{def:nbhd} and Definition~\ref{def:nbhd_in_dir},
it follows that \Nb{p}{C} is the union of \Nb{p}{x_i} for all the
variables of the clause $C$.

\begin{definition}
Let $X =\s{x_1,\dots,x_n}$ be a set of Boolean variables.  Let
$P=B_1\times\dots\times B_n$ be a cube of \spc and $B_i$ be equal to
\s{0} or \s{1}.  Denote by \bm{\mi{Nbhd}(P,x_i)} the union of
\Nb{p}{x_i} for all the points \pnt{p} of $P$. It is not hard to see
that $\mi{Nbhd}(P,x_i)$ is the cube obtained from $P$ by replacing
$B_i$ with the set $\s{0,1} \setminus B_i$. We will call
$\mi{Nbhd}(P,x_i)$ the \tb{1-neighborhood cube} of $P$ in direction
$x_i$.
\end{definition}

\begin{definition}
We will say that a \tb{cube} $P$ \tb{falsifies} a clause $C$ if $P
\subseteq \mi{Unsat}(C)$. (Obviously, in this case, every point of $P$
falsifies $C$.)
\end{definition}

\begin{definition}
Let $X =\s{x_1,\dots, x_n}$ be a set of Boolean variables.  Let $P$ be
a cube of \spc and $C$ be a clause falsified by $P$. Denote by
\bm{\mi{Nbhd}(P,C)} the set of 1-neighborhood cubes $\mi{Nbhd}(P,x_i)$
in every direction $x_i \in \V{C}$.
\end{definition}

Note that cubes satisfy the two requirements to clusters specified in
Remark~\ref{rem:kolmog_clust} of Section~\ref{sec:clusters}. On one
hand, the set of points contained in a cube $P$ can be succinctly
described. On the other hand, if a clause $C$ is falsified by $P$, the
1-neighborhood $\mi{Nbhd}(P,C)$ is the union of a small number of
cubes. So it can be easily computed.

\begin{definition}
Clauses $C',C''$ are called \tb{resolvable} on a variable $x$ if they
have the opposite literals of only one variable and this variable is
$x$. The clause $C$ is said to be obtained by \tb{resolution} of
$C',C''$ on $x$ if it consists of all the literals of $C',C''$ but
those of $x$. The clause $C$ is also called the \tb{resolvent} of
$C',C''$ on $x$.
\end{definition}

\begin{definition}
\label{def:cube_as_conj}
Let $P=B_1\times..\times B_n$ be a cube of \spc where $X\!=\!
\s{x_1,..,x_n}$. We will \tb{represent} $P$ as a \tb{conjunction} of
literals where the $i$-th literal component of $P$ corresponds to the
literal $l(x_i)$ of this conjunction and vice versa. Namely,
\begin{itemize}
\item $B_i = \s{0} \Leftrightarrow l(x_i) = \overline{x_i}$.
 \item $B_i = \s{1} \Leftrightarrow l(x_i) = x_i$.  
\end{itemize}
For every assignment $\pnt{p} \in P$ this conjunction evaluates to 1
and vice versa.
\end{definition}

\begin{example}
Let $P = \s{0,1} \times \s{0} \times \s{0,1} \times {1}$ be a cube of
\spc where $X=\s{x_1,x_2,x_3,x_4}$. Then $P$ can be specified by the
conjunction $\overline{x}_2 \wedge x_4$. For the sake of simplicity we
\tb{will omit} the sign $'\wedge'$. Then the cube $P$ is
\tb{specified} as $\overline{x}_2\,x_4$.
\end{example}

\begin{definition}
\label{def:merge}
Let $C',C''$ be two clauses resolvable on a variable $x_i \in
X$. Let $P',P''$ be two cubes of \spc that falsify $C'$ and $C''$
respectively. (This implies that the $i$-th component of $P'$ and
$P''$ is \s{b} and \s{\overline{b}} respectively where $b \in \s{0,1}$.)
Let $C$ be the resolvent of $C'$ and $C''$ on $x_i$.  A cube $P$ is
said to be obtained by \tb{merging} $P',P''$ on $x_i$ if
\begin{itemize}
 \item $P' \subseteq P$ and $P'' \subseteq P$
 \item $P$ falsifies $C$
\end{itemize}
\end{definition}

\begin{example}
Let $X\!=\!\s{x_1,\dots,x_8}$ and $C' = x_1 \vee x_3 \vee x_7$ and
$C'' = \overline{x}_1 \vee x_7$. Let $P' =
\overline{x}_1\,\overline{x}_3\,x_5\,\overline{x}_7\,x_8$ and $P'' =
x_1\,\overline{x}_3\,x_5\,\overline{x}_7$ be cubes of \spc falsifying
the clauses $C'$ and $C''$ respectively. Let $P =
\overline{x}_3\,x_5\,\overline{x}_7$. Note that $P' \subseteq P$ and
$P'' \subseteq P$. Besides, $P$ falsifies the resolvent $C = x_3 \vee
x_7$ of $C'$ and $C''$.  So $P$ can be viewed as obtained by merging
$P'$ and $P''$ on $x_1$. Note that, in general, a cube satisfying the
two conditions of Definition~\ref{def:merge} \tb{is not unique}. For
instance, the cube $P = \overline{x}_3\,\overline{x}_7$ satisfies
Definition~\ref{def:merge} as well.
\end{example}
\begin{definition}
\label{def:cov}
Let $P$ be a cube and $A$ be a set of cubes \s{P_1,\dots,P_k}.  We
will say that $P$ is \tb{covered} by $A$\linebreak if $P \subseteq
\mi{Union}(A)$ where \bm{\mi{Union}(A)} $= P_1 \cup \dots \cup
P_k$. That is the set of points specified by $P$ is a subset of the
set of points specified by $A$.
\end{definition}
%
%
\subsection{An example of how \Gs operates}
\label{ssec:example}
In this subsection, we give an example of how \Gs operates.  Consider
the formula $F(X)\!=\!C_1 \wedge \dots \wedge\!C_5$ where $C_1\!=\!
x_2 \vee x_3, C_2 = x_1 \vee\, \overline{x}_2, C_3 = \overline{x}_1
\vee\, \overline{x}_2 \vee x_3,\linebreak C_4 = \overline{x}_3 \vee
x_4, C_5 = \overline{x}_3 \vee \overline{x}_4$, $X =
\s{x_1,x_2,x_3,x_4}$.

%
%
\vspace{4pt}
\setlength{\intextsep}{4pt}
\setlength{\textfloatsep}{4pt}
\begin{figure}[h!]
\centering
\small
\begin{tabbing}
aaa\=bb\=cc\= dd\= \kill
initialize: \\
\scriptsize{1}\>  $P_1 = \overline{x}_2\,\overline{x}_3$, $g(P_1) = C_1=x_2 \vee x_3$ \\
\scriptsize{2}\>  $\mi{Body} = \emptyset, \mi{Boundary} = \s{P_1}$ \\[2pt]
compute $\mi{Nbhd}(P_1,C_1)$:\\
\scriptsize{3}\> $P_2 = \mi{Nbhd}(P_1,x_2) = x_2\,\overline{x}_3$ \\
\scriptsize{4}\> $P_3 = \mi{Nbhd}(P_1,x_3) = \overline{x}_2\,x_3$ \\
\scriptsize{5}\> $\mi{Body} = \s{P_1}, \mi{Boundary} = \s{P_2,P_3}$ \\[2pt]
splitting $P_2$ on $x_1$ \\
\scriptsize{6}\>$P'_2 = \overline{x}_1\,x_2\,\overline{x}_3$, $g(P'_2) = C_2= x_1 \vee\, \overline{x}_2$ \\
\scriptsize{7}\>$P''_2 = x_1\,x_2\,\overline{x}_3$, $g(P''_2) = C_3= \overline{x}_1
\vee\, \overline{x}_2 \vee x_3$\\
\scriptsize{8}\>$\mi{Body} = \s{P_1}, Boundary = \s{P'_2,P''_2,P_3}$ \\[2pt]
merging cubes $P'_2,P''_2$: \\
\scriptsize{9}\>$\mi{Merge}(P'_2,P''_2,x_1) = P_2 = x_2\,\overline{x}_3$ \\
\scriptsize{10}\>$C_6 = \mi{Res}(C_2,C_3,x_1) = \overline{x}_2 \vee x_3$ \\
\scriptsize{11}\>$\mi{Body} = \s{P_1}, \mi{Boundary} = \s{P_2,P_3}$,\\
\scriptsize{12}\>$F= F \wedge C_6$, $g(P_2) = C_6$\\[3pt]
compute $\mi{Nbhd}(P_2,C_6)$:\\
\scriptsize{13}\> $\mi{Nbhd}(P_2,x_2) =  \overline{x}_2\,\overline{x}_3 = P_1$ and $P_1 \in \mi{Body}$ \\
\scriptsize{14}\> $P_4 = \mi{Nbhd}(P_2,x_3) =  x_2\,x_3$ \\
\scriptsize{15}\>$\mi{Body} = \s{P_1,P_2}, \mi{Boundary} = \s{P_3,P_4}$,\\
~~~~~~...................................
\end{tabbing}
\vspace{-15pt}
\caption{A fragment of an execution trace of \Gs}
\label{fig:fragment}
\end{figure}

A fragment of the execution trace of \Gs applied to $F$ is shown in
Fig.~\ref{fig:fragment}. (Appendix~\ref{app:example} provides the
\ti{complete} trace where building an SSC for $F$ is finished proving
the latter unsatisfiable.)  Like \gs, \Gs maintains the sets \ti{Body}
and \ti{Boundary}. The difference is that these sets consist of
\tb{cubes} rather than points. \Gs starts with picking an initial cube
of the set \ti{Boundary} (line 1). Assume that \Gs picks the cube
$P_1$ equal to $\overline{x}_2\,\overline{x}_3$ as the initial
cube. $P_1$ falsifies the clause $C_1 = x_2 \vee x_3$.  So, \Gs adds
$g(P_1) := C_1$ to the definition of the transport function $g$.  At
this point, $\mi{Body} = \emptyset$ and $\mi{Boundary} = \s{P_1}$
(line 2).

Then \Gs computes $\mi{Nbhd}(P_1,C_1)$ \ie the neighborhood of $P_1$
with respect to $C_1$ (lines 3-5). It consists of the cubes $P_2 =
x_2\,\overline{x}_3$ and $P_3 = \overline{x}_2\,x_3$ obtained from
$P_1$ by negating the literals of $x_2$ and $x_3$ respectively. $P_1$
is moved from \ti{Boundary} to \ti{Body} and $P_2$, $P_3$ are added to
\ti{Boundary}.

Assume \Gs picks $P_2$ to replace it in \ti{Boundary} with the
1-neighborhood cubes.  Since $P_2$ does not falsify any clause of $F$,
\Gs cannot immediately compute the 1-neighborhood of $P_2$.  So, \Gs
splits it on $x_1$ replacing $P_2$ with cubes $P'_2 =
\overline{x}_1\,x_2\,\overline{x}_3$ and $P''_2 =
x_1\,x_2\,\overline{x}_3$ (lines 6-8). Note that $P'_2$ falsifies
$C_2$ and $P''_2$ falsifies $C_3$.

To compute the 1-neighborhood of $P_2$, \Gs merges $P'_2$ and $P''_2$
on $x_1$.  This merging reproduces $P_2$ and generates a new clause
falsified by it (lines 9-12). It is not hard to show that $P_2$ indeed
satisfies Definition~\ref{def:merge}. First, $P'_2 \subseteq P_2$ and
$P''_2 \subseteq P$.  Second, $P_2$ falsifies the new clause $C_6$
obtained by resolving $C_2$ and $C_3$ (falsified by $P'_2$ and $P''_2$
respectively).

Now \Gs is able to compute $\mi{Nbhd}(P_2,C_6)$ (lines 13-15). The
1-neighborhood cube in direction $x_2$ is covered by \ti{Body} since
this cube equals to $P_1$ and $P_1 \in \mi{Body}$. So, it is not added
to \ti{Boundary}. On the other hand, $\mi{Nbhd}(P_2,x_3)$ is a new
cube $P_4$ that is added to \ti{Boundary}.  As we mentioned above, the
rest of the execution trace is given in Appendix~\ref{app:example}.

\subsection{Procedure for building an SSC using cubes as clusters}
\label{ssec:SSC_with_cubes}
In this subsection, we present the pseudocode of \Gs
(Figure~\ref{fig:ssc}). \Gs accepts a formula $F$ and returns a
satisfying assignment or an SSC proving $F$ unsatisfiable. As we
mentioned above, like \gs, \Gs maintains sets \ti{Boundary}
and \ti{Body}. Here \ti{Boundary} (respectively \ti{Body}) is the set
of cubes whose 1-neighborhood cubes have not been generated yet
(respectively are already added to \ti{Boundary}).

%
%
\setlength{\intextsep}{4pt}
\setlength{\textfloatsep}{4pt}
\begin{figure}[h!]
\centering
\small
\vspace{5pt}
\begin{tabbing}
aaa\=bb\=cc\= dd\= \kill
$\Gs(F)$ \{ \\
\scriptsize{1}\> $P_{init} := \mi{gen\_cube}(F)$ \\
\scriptsize{2}\> $\mi{Body} =\emptyset, \mi{Boundary} = \s{P_{init}}$\\
\scriptsize{3}\> while ($\mi{Boundary} \neq \emptyset$) \{\\
\scriptsize{4}\Tt $P := \mi{pick\_next\_cube}(\mi{Boundary})$\\
\scriptsize{5}\Tt $\mi{Boundary} := \mi{Boundary} \setminus \s{P}$\\
\scriptsize{6}\Tt $H = \mi{falsified\_clauses}(F,P)$\\
\scriptsize{7}\Tt if ($H = \emptyset$) \{ \\
\scriptsize{8}\ttt if $(\mi{satisfies}(P,F))$ // every $\pnt{p} \in P$ \tb{satisfies} $F$\\
\scriptsize{9}\tttt return($P,\emptyset$) \\
\scriptsize{10}\ttt  $x := \mi{pick\_var}(P,F)$ \\
\scriptsize{11}\ttt $(P',P'') := \mi{split\_cube}(P,x)$\\
\scriptsize{12}\ttt  $\mi{Boundary} := \mi{Boundary} \cup \mi{uncov}(P',P'',\mi{Total})$ \\
\scriptsize{13}\ttt continue \}\\
\scriptsize{14}\Tt $(C',P',\mi{Merged}) := \ti{merge\_cubes}(P,\mi{Boundary},F)$ \\
\scriptsize{15}\Tt if ($|\mi{Merged}| > 1$) \{ // merging is successful\\
\scriptsize{16}\ttt  $\mi{Boundary} := (\mi{Boundary} \setminus \mi{Merged}) \cup \s{P'}$ \\
\scriptsize{17}\ttt  $F := F \wedge C'$ \\
\scriptsize{18}\ttt  continue \} \\
\scriptsize{19}\Tt $C := \mi{pick\_clause}(H)$ // $g(P) := C$ \\
\scriptsize{20}\Tt $\mi{NewCubes} :=  \mi{uncov}(\mi{Nbhd}(P,C),\mi{Total})$\\
\scriptsize{21}\Tt $\mi{Boundary} := \mi{Boundary} \cup \mi{NewCubes}$ \\
\scriptsize{22}\Tt $\mi{Body} := \mi{Body} \cup \s{P}$\}\\
\scriptsize{23}\> return($\mi{nil},\mi{Body}$) // $\mi{Body}$ is an SSC $=>$ $F$ is \tb{unsatisfiable}\\
\end{tabbing}
\vspace{-15pt}
\caption{Generation of SSC}
\label{fig:ssc}
\end{figure}

\Gs starts with producing a cube $P_{init}$ (line 1) to initialize the
set $\mi{Boundary}$ (line 2). $\mi{Body}$ is initially empty. An SSC
is built in a \ti{while} loop (lines 3-22). First, a cube $P$ is
picked and removed from $\mi{Boundary}$ (lines 4-5). Then the set $H$
of clauses falsified by $P$ is formed \ie for every clause $C$ of $H$,
it is true that $P \subseteq \mi{Unsat}(C)$ (line 6).

After that, \Gs checks if $H$ is empty (line 7). If so, there are the
two possibilities below. First, for every clause $C$ of $F$ it is true
that $\mi{Unsat}(C) \cap P = \emptyset$. This means that every point
$\pnt{p} \in P$ satisfies $F$ (line 8). So, \Gs returns $P$ and an
empty SSC (line 9). The second possibility is that there are clauses
$C$ of $F$ such that $\mi{Unsat}(C) \cap P \neq \emptyset$, but none
of them is falsified by the cube $P$. In that case, \Gs splits the
cube $P$ (lines 10-11) on a variable $x$ into cubes $P'$ and
$P''$. Both cubes are tested by the function \ti{uncov}, if they are
covered by \ti{Total} (see Definition~\ref{def:cov}) \ie whether $P'$
or $P''$ is a subset of $\mi{Union}(\mi{Total})$.  Here \bm{\mi{Total}
  = \mi{Body} \cup \mi{Boundary}} and \bm{\mi{Union}(\mi{Total})} is
the union of the cubes of \ti{Total}. If $P'$ or $P''$ is not covered
by \ti{Total}, it is added to $\mi{Boundary}$ (line 12). Checking if
$P'$ or $P''$ is covered can be done by a regular SAT-solver (see the
discussion of Subsection~\ref{ssec:improvements}).

If $H$ is not empty, \Gs invokes the procedure $\mi{merge\_cubes}$
(line 14). It tries to merge the cube $P$ with other cubes of
$\mi{Boundary}$ to reduce the size of the latter.  To this end,
$\mi{merge\_cubes}$ applies multiple merge operations described by
Definition~\ref{def:merge}. It returns a) the subset $\mi{Merged}$ of
$\mi{Boundary}$ consisting of the merged cubes including the cube $P$
and b) a cube $P'$ obtained by merging the cubes of $\mi{Merged}$ and
c) a new clause $C'$ falsified by $P'$ that is obtained by resolving
clauses of $F$ falsified by cubes of $\mi{Merged}$. If
$\mi{merge\_cubes}$ succeeds, $|\mi{Merged}| > 1$. Then the merged
cubes are removed from $\mi{Boundary}$, $P'$ is added to
$\mi{Boundary}$ (line 16) and $C'$ is added to $F$ (line 17). Then a
new iteration begins.

If $\mi{merge\_cubes}$ fails, \Gs picks a clause $C$ of $H$ (line 19)
and forms the set of cubes $\mi{Nbhd}(P,C)$. The function $\mi{uncov}$
discards every cube of $\mi{Nbhd}(P,C)$ covered by \ti{Total} (line
20). The cubes of $\mi{Nbhd}(P,C)$ that have not been discarded are
added to $\mi{Boundary}$ (line 21).  Finally, $P$ is added to
$\mi{Body}$ and a new iteration of the loop begins.

If \ti{Boundary} is empty, then \ti{Body} is an SSC.  \Gs returns the
latter as a proof that $F$ is unsatisfiable (line 23).

\section{Discussion Of \Gs}
\label{sec:discussion}
In this section, we discuss \Gs.  In Subsection~\ref{ssec:snd_cmpl} we
give propositions stating that \Gs is sound and complete.
Subsection~\ref{ssec:improvements} discusses potential improvements of
\Gs. In Subsection~\ref{ssec:par_sat}, we argue that \Gs facilitates
parallel solving.

\subsection{\Gs is sound and complete}
\label{ssec:snd_cmpl}
\Gs terminates when it builds a cube $P$ satisfying every clause of
$F$ (line 9) or when the set \ti{Boundary} is empty (line 23). The
latter means that the set of points \ti{Union}(\ti{Body}) forms an
SSP. In the first case $F$ is correctly reported as satisfiable and in
the second case it is properly identified as unsatisfiable. So, the
proposition below holds. (As mentioned earlier, proofs of the new
propositions are given in the appendix.)
\begin{proposition}
  \label{prop:sound}
If \Gs terminates, it returns the correct answer \ie \Gs
is \ti{sound}.
\end{proposition}

One can also show that \Gs is complete. Here is a high-level
explanation why. The function $\xi = |\mi{Union}(\mi{Body})| + |F|$
cannot decrease its value during the operation of \Gs. That is $\xi$
either grows or keeps its value unchanged. For instance, if \Gs moves
a cube from \ti{Boundary} to \ti{Body} the value of $\xi$ increases.
The same occurs after a new clause is generated and added to $F$ when
\Gs merges cubes of \ti{Boundary}.  In the proof of completeness of
\Gs we show that the number of steps where $\xi$ \ti{preserves} its
value is finite. This observation and the fact that the range of $\xi$
is finite too implies that \Gs always terminates. Hence, the
proposition below holds.
\begin{proposition}
  \label{prop:complete}
\Gs terminates for every CNF formula \ie \Gs is \ti{complete}.
\end{proposition}
%
%
\subsection{Improvements to \Gs}
\label{ssec:improvements}
The main flaw of the version of \Gs described in Fig.~\ref{fig:ssc} is
as follows. Let $P^*$ be either a cube obtained by splitting a cube of
\ti{Boundary} or a 1-neighborhood cube of a cube of \ti{Boundary}. To
find out if $P^*$ needs to be added to \ti{Boundary}, \Gs checks if
\ti{Total} covers $P^*$ (lines 12 and 20). That is, if $P^*$ is a
subset of $\mi{Union}(\mi{Body} \cup \mi{Boundary})$.  This check can
be performed by an ``auxiliary'' SAT solver based on conflict clause
learning~\cite{grasp,chaff}. The problem however is that such a check
can be computationally hard.

There are at least two methods to address this problem. The first
method is to make the auxiliary SAT solving easy by checking only if
$P^*$ is covered by a small subset of $\mi{Body} \cup
\mi{Boundary}$. For instance, this subset may include only cubes
sharing literal components with $P^*$.  The other method is to combine
regular SAT solving based on conflict clause learning and computing an
SSC~\cite{cdcl+ssc}.  This method avoids auxiliary SAT solving at the
expense of building an SSC specifying a larger SSP. Importantly, even
if \Gs builds an SSC specifying the trivial SSP equal to \spc, the SAT
algorithm remains ``local'' since it does not produce a ``global''
certificate of unsatisfiability \ie an empty clause. So, in a sense,
the size of the SSP specified by SSC does not matter.
%
%
\subsection{Parallel SAT computing}
\label{ssec:par_sat}
Creating efficient algorithms of parallel SAT solving is a tall
order~\cite{par_SAT}. One of the main problems here is that the SAT
procedures used in practice prove unsatisfiability by resolution.  A
resolution proof can be viewed as \tb{global} in the sense that a) it
has a global goal (derivation of an empty clause) and b) different
parts of the proof strongly depend on each other. This makes
resolution procedures hard to parallelize.

On the other hand, an SSP can be viewed as a \tb{local} proof. Given a
formula $F$, one just needs to find a set of points $P$ and a
transport function $g:P \mapsto F$ such that a \ti{local} property
holds. Namely, for every point $\pnt{p} \in P$, the relation
$\mi{Nbhd}(\pnt{p},C) \subseteq P$ holds where $C = g(\pnt{p})$.  Note
that in contrast to a resolution proof, generation of an SSP does not
have a global goal. So, arguably, building an SSP is easier to
parallelize.  Importantly, constructing an SSC produces a local proof
as well since one simply builds an SSP \ti{implicitly} (via clusters
of points). So, one can argue that generating an SSC facilitates
parallel computing too.

\section{Some Background}
\label{sec:background}
In this section, we briefly discuss the relation of SSPs and SAT
algorithms based on local search (Subsection~\ref{ssec:loc_search})
and, in particular, the derandomized version of the Sh\"oning
procedure (Subsection~\ref{ssec:derandom}). Besides, we relate SSCs
with two ``local'' proof systems we introduced earlier
(Subsection~\ref{ssec:relation}).

\subsection{Local search procedures}
\label{ssec:loc_search}
SAT algorithms based on local search have been a subject of study for
a long time. First, local search was applied only to satisfiable
formulas. Papadimitriou showed~\cite{rand_walk} that a very simple
stochastic local search procedure finds a satisfying assignment of a
2-CNF formula in polynomial time. Then, a few practical SAT-algorithms
based on stochastic local search were developed and successfully
applied to more general classes of satisfiable CNF
formulas~\cite{noise}. In~\cite{prob_alg} a new powerful stochastic
algorithm for solving satisfiable CNF formulas was introduced by
Sh\"oning. Later, a derandomized version of that algorithm was
developed that achieved the best known upper bound on complexity of
solving $k$-SAT~\cite{derandom}. We will refer to this procedure as
\bm{\sd} where \ti{Schn} stands for \ti{Sh\"oning} and \ti{der} for
\ti{derandomized}. Importantly, \sd can be applied to both satisfiable
and unsatisfiable CNF formulas.

On the one hand, SSPs can be related to local search algorithms. In
particular, the \gs procedure recalled in
Subsection~\ref{ssec:bld_ssp} looks similar to \sd (see the the
next subsection). On the other hand, the definition of an SSP is
algorithm independent, which makes SSPs a very appealing object of
study and separates them from the local search algorithms. This
distinction becomes more conspicuous in this report where we consider
the notion of a stable set of clusters. For example, the \Gs procedure
where clusters are cubes of points (see Section~\ref{sec:cubes}) does
not look like a local search procedure at all.

%
%
%
\subsection{\sd and SSPs}
\label{ssec:derandom}
In this subsection, we compare \sd and \gs computing an SSP point by
point. Let $F$ be a formula to check for satisfiability.  \sd consists
of two steps. First, \sd computes a set of Boolean balls covering the
entire search space \spc where $X = \V{F}$. A \tb{Boolean ball} with a
center \pnt{p} and radius $r$ is the set of all points \pnt{p'} such
that $0 \leq \mi{distance}(\pnt{p},\pnt{p'}) \leq r$. (Here,
\ti{distance} specifies the \ti{Boolean distance}.) Second, for every
Boolean ball, \sd runs a procedure \Sr that checks if this ball
contains a satisfying assignment.

\gs can be viewed~\cite{priv_comm2} as a version of \sd covering the
space \spc with balls of radius $r=1$. Indeed, given a point \pnt{p}
and a clause $C$ falsified by \pnt{p}, checking the 1-neighborhood
\Nb{p}{C} mimics what the call $\sr(F,\pnt{p},1)$ does.  The main
difference between \gs and \sd is that, in contrast to the latter, the
Boolean balls of the former \ti{talk to each other}. This allows \gs
to claim $F$ to be unsatisfiable as soon as the set of visited balls
becomes stable. Importantly, the idea of reaching the stability of
talking Boolean balls can be extended to a huge variety of clusters of
points (see Remark~\ref{rem:kolmog_clust}).  Moreover, as we mentioned
earlier, one can extend the notion of stability to multi-level
clusters (e.g. clusters of clusters of points). The clustering of
points serves here two purposes. First, it allows one to speed up SAT
solving. Second, it facilitates exploiting the structure of the
formula at hand by making clusters formula-specific.

\subsection{Relation to proof systems NE and NER}
\label{ssec:relation}
In~\cite{provinglocally}, we introduced two ``local'' proof systems,
\ti{NE} and \ti{NER}. These proof systems are based on the fact that
if a CNF formula $F$ is satisfiable, there always exists a satisfying
assignment \pnt{p} that satisfies only one literal of some clause $C$
of $F$. $($In terms of this report, \pnt{p} is located in
$\mi{Nbhd}(\mi{Unsat}(C),C)$ \ie in the 1-neighborhood of $C$ with
respect to the cube $\mi{Unsat}(C))$. The idea of either proof system
is to explore the 1-neighborhood of all the clauses of $F$.  The
difference between \ti{NE} and \ti{NER} is that the latter allows one
to use resolution to generate new clauses.

Let $F$ be equal to $C_1\wedge \dots \wedge C_k$. One can show that
\Gs generates proofs similar to those of \ti{NE} and \ti{NER} if the
set \ti{Boundary} is initialized with the cubes\linebreak
$\mi{Unsat}(C_i),i=1,\dots,k$. More precisely, \ti{NE} is similar to
\Gs without the option of merging cubes of \ti{Boundary} (and thus
without the option of resolving clauses) and \ti{NER} is similar to
\Gs if cube merging is allowed. The main flaw of \ti{NE} and \ti{NER}
is that if $F$ contains a small unsatisfiable subset of clauses, a
proof in \ti{NE} and \ti{NER} still involves \ti{all} clauses of
$F$. (The notion of an SSP was actually designed to address this flaw
of \ti{NE} and \ti{NER}.) On the other hand, in the case above, \Gs
can produce an SSC that involves only a small fraction of clauses of
$F$.

\section{Conclusions}
\label{sec:conclusions}
Earlier we introduced the notion of a stable set of points (SSP) for a
CNF formula $F$.  (Here, a point is a complete assignment to the
variables of $F$.)  A CNF formula $F$ is satisfiable if and only if it
has a stable set of points.  In this paper we present the notion of a
stable set of clusters (SSC) of points.  The main goal of using SSCs
is to speed up the construction of an SSP for $F$ by processing many
points at once.  We give two methods of computing SSCs. In the first
method, clusters are specified by equivalence classes describing
permutational symmetries of $F$. This method is an example of an
algebraic approach to SAT solving. Importantly, one can extend the
notion of a stable set of two-level clusters (\ie clusters of points)
to multi-level ones (\eg clusters of clusters of points). In the
second method, clusters are represented by cubes. In contrast to the
first method that can be applied only to formulas with permutational
symmetries, this method can be used for any CNF formulas. In addition
to direct SAT solving, the second method can be employed for better
understanding and improving the performance of existing SAT algorithms
based on resolution.


\bibliographystyle{plain}
\bibliography{short_sat,local}
\appendices
\section{Proofs Of Propositions}
\label{app:proofs}
\setcounter{proposition}{1}
\begin{proposition}
Let $F$ be a CNF formula and $P_1,\dots,P_k$ be a stable set of
clusters with respect to $F$ and transport functions $g_1,..,
g_k$. Then $P_1 \cup \dots \cup P_k$ is an SSP and so $F$ is
unsatisfiable.
\end{proposition}
\begin{mmproof}
Denote by $P$ the set $P_1 \cup \dots \cup P_k$. Let $g$ be a
transport function such that for every $\pnt{p} \in Z(F)$, it is true
that\linebreak $g(\pnt{p}) = C$, where $C \in F$ and $C=g_i(\pnt{p})$,
$1 \leq i \leq k$. In other words, $g$ assigns to \pnt{p} the same
clause that is assigned to \pnt{p} by a function $g_i$ (picked
arbitrarily from $g_1,\dots,g_k$). Then $P$ is an SSP with respect to
$F$ and the transport function $g$. Indeed, let \pnt{p} be a point of
$P$ and $g_i$ be a transport function such that
$g(\pnt{p})=g_i(\pnt{p})=C$. Since \s{P_1,\dots,P_k} is an SSC, then
$\mi{Nbhd}(P_i,g_i) \subseteq P$. Hence $\Nb{p}{g_i(\pnt{p})}
\subseteq P$ and so $\Nb{p}{g(\pnt{p})} \subseteq P$.
\end{mmproof}

\setcounter{proposition}{4}
\begin{proposition}
If \Gs terminates, it returns the correct answer \ie \Gs
is \ti{sound}.
\end{proposition}
\begin{mmproof}
Let $F$ be a CNF formula to test for satisfiability.
\Gs returns the answer \ti{satisfiable} (line 9 of Fig.~\ref{fig:ssc}) only
if an assignment satisfying $F$ is found. So the
answer \ti{satisfiable} is always correct.

Now we show that if \Gs reports that $F$ is unsatisfiable (line 23),
the set \ti{Body} is an SSC for $F$.  So the answer \ti{unsatisfiable}
is also always correct. Let $P$ be a cube of \ti{Body} and $C$ be the
clause assigned to $P$ by the transport function $g$ \ie $g(P) = C$
and $P$ falsifies $C$.  Originally, $P$ appears in the
set \ti{Boundary} and is moved to \ti{Body} only when the cubes of
$\mi{Nbhd}(P,C)$ are generated (lines 19-22).

Let $P'$ be an arbitrary cube of $\mi{Nbhd}(P,C)$.  Let us show that
$P'$ will be covered by the final set \ti{Body} and so the latter is
an SSC for $F$.  Indeed, if $P'$ is not covered by the current set
$\mi{Body} \cup \mi{Boundary}$, it is added to \ti{Body} and hence it
will be present in the final set \ti{Body}. If $P'$ is covered by the
set $\mi{Body}\,\cup\,\mi{Boundary}$, there are two cases to consider.
If $P'$ is covered by \ti{Body} alone, this means that every point of
$P'$ is already in \ti{Union}(\ti{Body}). So $P'$ will be covered by
the final set \ti{Body}. If $P'$ is \ti{not} covered by \ti{Body},
then some points of $P'$ are present only in the current
set \ti{Boundary}. Since eventually \ti{Boundary} becomes empty, the
cubes containing those points of $P'$ will be moved to \ti{Body}. So,
again, $P'$ will be covered by the final set
\ti{Body}.

\end{mmproof}

\begin{proposition}
\Gs terminates for every CNF formula \ie \Gs is \ti{complete}.
\end{proposition}
\begin{mmproof}
Assume the contrary \ie
\Gs does not terminate on a CNF formula $F$.  Consider
the function\linebreak $\xi = |\mi{Union}(\mi{Body})| + |F|$. Note
that $\xi$ cannot reduce its value. That is, in every iteration of the
loop of \Gs, this value either stays the same or increases due to the
growth of $|\mi{Union}(\mi{Body})|$ and/or $|F|$. Since the maximum
value of $\xi$ is $2^n+3^n$ (where $n=|\V{F}|$), \Gs can have only a
finite set of iterations of the loop in which $\xi$ grows. Since, by
our assumption, \Gs does not terminate, there exists an infinite
sequence of iterations in which $\xi$ preserves its value. Let us show
that this is not the case.

The value of $\xi$ does not change only when the cube $P$ picked from
\ti{Boundary} is split or when every cube of $\mi{Nbhd}(P,C)$ is covered by \ti{Total}.
(Recall that $\mi{Total} = \mi{Body}\,\cup\,\mi{Boundary}$.) In the first case,
$P$ is replaced in \ti{Boundary} with two cubes of
a smaller size. In the second case, $P$ is just removed
from \ti{Boundary}. The number of splits performed on a cube and its
descendants is bound by $2^n$. So, the total number of splits of the
cubes of \ti{Boundary} is limited by $3^n*2^n$ where $3^n$ is the
upper bound on $|\mi{Boundary}|$ (because $3^n$ is the total number of
different cubes of $n$ variables).  The total number of iterations
that remove a cube from \ti{Boundary} without adding it to \ti{Body}
is also limited by $3^n$. So, the total number of iterations that do
not change the value of $\xi$ is limited by $(2^n+1)*3^n$. Before
this limit is exceeded, an event below takes place.
\begin{itemize}
\item A cube obtained by splitting satisfies all clauses of $F$ and \Gs terminates.
\item A cube  of $\mi{Nbhd}(P,C)$ that is not covered by \ti{Total} is added to
\ti{Body} thus increasing the value of  $\xi$.
\item A new clause is produced and added to $F$ when merging cubes of \ti{Boundary}, which increases the value of $\xi$.
\item \ti{Boundary} becomes empty and \Gs terminates reporting that \ti{Body}
is an SSC and thus $F$ is unsatisfiable.
\end{itemize}
In every case above, \Gs either terminates or the value of $\xi$
increases.  So, \Gs cannot have an infinite sequence of iterations
where $\xi$ does not change its value. Hence, \Gs always terminates.
\end{mmproof}
\raggedbottom

\section{An example of how \Gs operates}
\label{app:example}
%
%
\begin{figure}[t!]
\small
\begin{tabbing}
aaa\=bb\=cc\= dd\= \kill
initialize: \\
\scriptsize{1}\>  $P_1 = \overline{x}_2\,\overline{x}_3$, $g(P_1) = C_1=x_2 \vee x_3$ \\
\scriptsize{2}\>  $\mi{Body} = \emptyset, \mi{Boundary} = \s{P_1}$ \\[2pt]
compute $\mi{Nbhd}(P_1,C_1)$:\\
\scriptsize{3}\> $P_2 = \mi{Nbhd}(P_1,x_2) = x_2\,\overline{x}_3$ \\
\scriptsize{4}\> $P_3 = \mi{Nbhd}(P_1,x_3) = \overline{x}_2\,x_3$ \\
\scriptsize{5}\> $\mi{Body} = \s{P_1}, \mi{Boundary} = \s{P_2,P_3}$ \\[2pt]
splitting $P_2$ on $x_1$ \\
\scriptsize{6}\>$P'_2 = \overline{x}_1\,x_2\,\overline{x}_3$, $g(P'_2) = C_2=x_1 \vee\, \overline{x}_2$ \\
\scriptsize{7}\>$P''_2 = x_1\,x_2\,\overline{x}_3$, $g(P''_2) = C_3= \overline{x}_1
\vee\, \overline{x}_2 \vee x_3$\\
\scriptsize{8}\>$\mi{Body} = \s{P_1}, Boundary = \s{P'_2,P''_2,P_3}$ \\[2pt]
merging cubes $P'_2,P''_2$: \\
\scriptsize{9}\>$\mi{Merge}(P'_2,P''_2,x_1) = P_2 = x_2\,\overline{x}_3$ \\
\scriptsize{10}\>$C_6 = \mi{Res}(C_2,C_3,x_1) = \overline{x}_2 \vee x_3$ \\
\scriptsize{11}\>$\mi{Body}=\s{P_1},\mi{Boundary}=\s{P_2,P_3}$\\
\scriptsize{12}\> $F=F \wedge C_6$, $g(P_2)= C_6$\\[3pt]
compute $\mi{Nbhd}(P_2,C_6)$:\\
\scriptsize{13}\> $\mi{Nbhd}(P_2,x_2) =  \overline{x}_2\,\overline{x}_3 = P_1$ and $P_1 \in \mi{Body}$ \\
\scriptsize{14}\> $P_4 = \mi{Nbhd}(P_2,x_3) =  x_2\,x_3$ \\
\scriptsize{15}\>$\mi{Body} = \s{P_1,P_2}, \mi{Boundary} = \s{P_3,P_4}$,\\
............................................\\
splitting $P_3$ on $x_4$: \\
\scriptsize{16}\>$P'_3 = \overline{x}_2\,x_3\,\overline{x}_4$, $g(P'_3) = C_4= \overline{x}_3 \vee
x_4$ \\
\scriptsize{17}\>$P''_3 = \overline{x}_2\,x_3\,x_4$, $g(P''_3) = C_5=\overline{x}_3 \vee \overline{x}_4$\\
\scriptsize{18}\>$\mi{Body} = \s{P_1,P_2}, Boundary = \s{P'_3,P''_3,P_4}$ \\[2pt]
merging cubes $P'_3,P''_3$: \\
\scriptsize{19}\>$P_3 = \mi{Merge}(P'_3,P''_3,x_4) = P_3 = \overline{x}_2\,x_3$ \\
\scriptsize{20}\>$C_7 = \mi{Res}(C_4,C_5,x_4) = \overline{x}_3$ \\
\scriptsize{21}\>$\mi{Body} = \s{P_1,P_2}, \mi{Boundary} = \s{P_3,P_4}$,\\
\scriptsize{22}\>$F= F \wedge C_7$, $g(P_3) = C_7$\\[3pt]
compute $\mi{Nbhd}(P_3,C_7)$:\\
\scriptsize{23}\> $\mi{Nbhd}(P_3,C_7) =  \overline{x}_2\,\overline{x}_3 = P_1$ and $P_1 \in \mi{Body}$ \\
\scriptsize{24}\>$\mi{Body} = \s{P_1,P_2,P_3}, \mi{Boundary} = \s{P_4}$,\\[2pt]
compute $\mi{Nbhd}(P_4,C_7)$:\\
\scriptsize{25}\> $\mi{Nbhd}(P_4,C_7) =  x_2\,\overline{x}_3 = P_2$ and $P_2 \in \mi{Body}$ \\
\scriptsize{26}\>$\mi{Body} = \s{P_1,P_2,P_3,P_4}, \mi{Boundary} = \emptyset$\\
finish: \\
\scriptsize{27}\> return \ti{Body}\\
\end{tabbing}
\vspace{-15pt}
\caption{Example of how \Gs operates}
\vspace{10pt}
\label{fig:full_trace}
\end{figure}

In this appendix, we complete the example of
Subsection~\ref{ssec:example}.  Namely, we describe the part of the
execution trace after the dotted line (lines 16-27,
Fig.~\ref{fig:full_trace}). At this point, $\mi{Body} = \s{P_1,P_2}$
and $\mi{Boundary} = \s{P_3,P_4}$.

\Gs picks $P_3 = \overline{x}_2\,x_3$
from \ti{Boundary} and splits it on variable $x_4$ (lines 16-18).  The
reason for splitting is that $P_3$ does not falsify any clause of $F$.
On the other hand, the cubes $P'_3$ and $P''_3$ produced by splitting
falsify clauses $C_4$ and $C_5$ respectively. $P_3$ is replaced
in \ti{Boundary} with $P'_3$ and $P''_3$.

Then \Gs merges cubes $P'_3$ and $P''_3$ to generate the cube $P_3$
again (lines 19-22). But now a new clause $C_7 = x_3$ is added to $F$
that is falsified by $P_3$. The clause $C_7$ is produced by resolving
clauses $C_4$ and $C_5$ falsified by $P'_3$ and $P''_3$.  The cubes
$P'_3,P''_3$ are replaced in \ti{Boundary} with $P_3$.

\Gs picks the cube $P_3$ again but now it is able to compute
its 1-neighborhood with respect to the clause $C_7$ (lines 23-24).
Since $C_7$ has only one literal, \nb{P_3}{C_7} consists of only one
cube. Since this cube equals $P_1$ that is already in \ti{Body}, $P_3$
is just moved from \ti{Boundary} to \ti{Body} without adding anything
to the former.

Finally, \Gs picks $P_4$, the last cube of \ti{Boundary}.  Since $P_4$
falsifies $C_7$, \Gs computes the 1-neighborhood of the former with
respect to the latter. This 1-neighborhood consists of the cube equal
to $P_2$ that is already in \ti{Body}. So,
\Gs just moves $P_4$ to \ti{Body}.

At this point the set \ti{Boundary} is empty.  This means that the
current set \ti{Body} is an SSC and $F$ is unsatisfiable.
\Gs returns \ti{Body} as a proof of unsatisfiability (line 27).

\end{document}